\documentclass[fleqn,usenatbib,useAMS]{mnras}

\usepackage{graphicx}	
\usepackage{amsmath}	
\usepackage{amssymb}	
\usepackage{multicol}        
\usepackage{multirow}
\usepackage{bm}		
\usepackage{pdflscape}	

\usepackage{xspace}
\usepackage[capitalize]{cleveref}
\usepackage[dvipsnames]{xcolor}
\usepackage{siunitx}

\usepackage{subcaption}
\usepackage[T1]{fontenc}
\usepackage{ae,aecompl}
\usepackage{newtxtext,newtxmath}
\usepackage{ulem}
\usepackage{soul}
\pubyear{2022}

\DeclareRobustCommand{\VAN}[3]{#2}
\let\VANthebibliography\thebibliography
\def\thebibliography{\DeclareRobustCommand{\VAN}[3]{##3}\VANthebibliography}

\begin{document}
\label{firstpage}
\pagerange{\pageref{firstpage}--\pageref{lastpage}}

\title[The mHz QPO of Cen X-3]{Detection of a quasi-periodic oscillation at $\sim$40 mHz in Cen X-3 with Insight-HXMT}

\author[Q. Liu et al.]{Q. Liu$^{1,2}$, W. Wang$^{1,2}$\thanks{E-mail: wangwei2017@whu.edu.cn}, X. Chen$^{1,2}$, W. Yang$^{1,2}$, F. J. Lu$^{3}$, L. M. Song$^{3}$, J. L. Qu$^{3}$, S. Zhang$^{3}$, and S. N. Zhang$^{3}$ \\
$^{1}$School of Physics and Technology, Wuhan University, Wuhan 430072, China\\
$^{2}$WHU-NAOC Joint Center for Astronomy, Wuhan University, Wuhan 430072, China\\
$^{3}$Key Laboratory of Particle Astrophysics, Institute of High Energy Physics, Chinese Academy of Sciences, Beijing 100049, China
}

\maketitle
\begin{abstract}
We investigated the quasi-periodic oscillation (QPO) features in the accretion-powered X-ray pulsar Cen X-3 observed by Insight-HXMT. For two observations in 2020 when Cen X-3 was in an extremely soft state, the power density spectrum revealed the presence of obvious QPO features at $\sim$40 mHz with an averaged fractional rms amplitude of $\sim$9\%. We study the mHz QPO frequency and rms amplitude over orbital phases, and find that the QPO frequency is $\sim$ 33--39  mHz at the orbital phase of 0.1-- 0.4, increasing to $\sim 37-43$ mHz in the orbital phase of 0.4--0.8, but has no strong dependence on X-ray intensity. We also carried out an energy-dependent QPO analysis, the rms amplitude of the mHz QPOs have a decreasing trend as the energy increases from 2 to 20 keV. In addition, the QPO time-lag analysis shows that the time delay is $\sim 20$ ms (a hard lag) in the range of $\sim$ 5--10 keV, and becomes negative (time lag of $-(20-70)$ ms) above $\sim 10$ keV. The different QPO theoretical models are summarized and discussed. In the end, we suggest that these energy-dependent timing features as well as the origin of mHz QPOs in Cen X-3 may be ascribed to an instability when the accretion disk is truncated near the corotation radius.
\end{abstract}


\section{Introduction}
\label{sec:introduction}

High-mass X-ray binaries (HMXBs) are comprised of a compact object (a neutron star or a black hole) and a high-mass companion star. The X-ray binaries are mostly powered by accreted matter from the donor’s powerful stellar wind or Roche lobe overflow. For accretion on to a highly magnetized neutron star, the accreted matter transfers onto the neutron star unaffected by the magnetic field lines until the Alfvén radius, where the pressure of the magnetic field balances the ram pressure of the infalling plasma. And through this position, the accretion flow is funneled to relatively small regions on the neutron star surface near the magnetic polar caps by the field and forms the hot spots \citep{1975A&A....42..311B}, where the X-ray emission occurs, with the power originating from the released kinetic energy \citep{1973ApJ...184..271L}.

Cen X-3 is an eclipsing (lasts $\sim$22 percent of the orbit) high-mass X-ray binary, first discovered by \cite{PhysRevLett.19.681}, and is the first confirmed X-ray pulsar \citep{1971ApJ...167L..67G}. The NS spin period is $\sim$4.8 s \citep{1971ApJ...167L..67G,2007AA...473..523V} with a spin-up trend \citep{1994AIPC..304..304F,1996ApJ...456..316T} and the NS orbits an optical companion O-type supergiant V779 Cen \citep{1972ApJ...172L..79S,1974ApJ...192L.135K,1979ApJ...229.1079H} per $\sim$2.087 d \citep{2015A&A...577A.130F} with an inclination angle of $\sim$79$^\circ$ \citep{2021MNRAS.501.5892S}. The neutron star of this system has a mass of $1.34_{-0.14}^{+0.16}$ $M_{\odot}$ \citep{2007AA...473..523V} with a strong magnetic field of (2.4–3.0) $\times$ $10^{12}$ G \citep{1998A&A...340L..55S} and the companion possesses a mass of $ \sim20.2_{-1.5}^{+1.8}$ $M_{\odot}$ \citep{2007AA...473..523V} and a radius of $\sim$12 $R_{\odot}$ \citep{Naik_2011}. The binary system is estimated to be at a distance of $\sim$8 kpc  \citep{1974ApJ...192L.135K}. However, using the energy-resolved dust-scattered X-ray halo, \cite{2009ApJ...691.1744T} estimated the distance to be 5.7 $\pm$ 1.5 kpc, which is consistent with the Gaia distance of 6.9 kpc \citep{2021MNRAS.507.3899V} using the galactic X-ray binary priors from \cite{2019MNRAS.489.3116A}.

As one of the brightest accretion X-ray sources in the Galaxy, The presence of an accretion disk of Cen X-3 has been claimed by \cite{1986A&A...154...77T} who applied a simple geometric model to interpret the observed optical light curves. The high luminosity of $\sim (4-5) \times 10^{37}$ erg\ s$^{-1}$ \citep{2008ApJ...675.1487S} also indicates the existence of an accretion wake in this system. The detection of quasi-periodic oscillations (usually linked to the disc) observed at $\sim$40 mHz \citep{1991PASJ...43L..43T,2008ApJ...685.1109R} is also consistent with the mode of accretion disc-fed by Roche lobe overflow. In addition, this neutron star's spectrum appears to have the cyclotron resonance scattering features (CRSF) around 30 keV \citep{1992ApJ...396..147N,1998A&A...340L..55S,2019A&A...622A..61S}, which make it one of the nine X-ray binary sources \citep{2021MNRAS.508.5578R} exhibiting a CRSF as well as a QPO. 

The long-term but aperiodic flux variability (\citealt{1983ApJ...273..709P}) on time scales (e.g., superorbital period) larger than its orbital period seen in Cen X-3 has been reported by \cite{2008MNRAS.387..439R} based on RXTE-ASM, which is attributed to the precession of the warped accretion disc. They note that the eclipse seen in the high state (orbit-averaged ASM count rate $>$ 18 in 1.5–12 keV defined by \citealt{2008MNRAS.387..439R}) of the superorbital period is sharp and it turns to more smooth and longer in the superorbital low state (orbit-averaged ASM count rate $<$ 2), which indicates a larger emission region in the low state. Investigation of X-ray spectrum of Cen X-3 shows the fluorescent emission of iron lines \citep{1996PASJ...48..425E}. 
The 6.4 keV iron line forms in cold material close to the NS \citep{1989PASJ...41....1N}, whereas the 6.67 keV and 6.97 keV iron emission lines probably originate in the highly photoionized accretion disc corona \citep{1989ApJ...341..955K}. 

In accretion-powered X-ray pulsars, QPOs are traditionally thought to be due to inhomogeneity in the inner accretion disk. 
For most of the accreting pulsars, the QPOs exhibit a transient nature. The presence of QPO features has been known both in transient and persistent X-ray sources \citep{2011BASI...39..429P} but QPOs seem to occur more in transient sources, such as EXO 2030+375 (200 mHz; \citealt{1989ApJ...346..906A}), A 0535+262 (50 mHz; \citealt{1996ApJ...459..288F}), 4U 0115+63 (10 mHz, \citealt{2021MNRAS.503.6045D}), and V0332+53 (51 mHz; \citealt{1994ApJ...436..871T}). Early X-ray studies of the persistent source Cen X-3 have shown QPOs at $\sim$35 mHz \citep{1991PASJ...43L..43T}. \cite{2008ApJ...685.1109R} reported the QPOs of the peak frequency around 40 and 90 mHz, with the QPO frequency having no dependence on X-ray intensity, then they claimed that the QPO generation mechanism in Cen X-3 is different from the beat frequency model or Keplerian frequency model. Therefore, the origin of the QPO related to the inhomogeneities in the inner disk for Cen X-3 is still questioned or in dispute \citep{2008ApJ...685.1109R}. A detailed temporal analysis of QPOs features could give new insight into the dynamic properties of these NS systems. So, we re-investigated the persistent source of Cen X-3 to study the features of QPOs and understand the physical origin. 

In this paper, we study the evolution of QPOs of Cen X-3 in relation to orbital phases using Insight-HXMT observations and find that QPO frequency drifts over the orbital phases. Then an energy-dependent QPO analysis and corresponding time lag analysis are further performed. The paper is organized as follows. In Section 2, we describe the HXMT observations and data reduction. The timing analysis of Cen X-3 and QPO results are presented in Section 3. In Section 4, we discuss the QPO features and possible explanations. Conclusions are summarized in Section 5.

\section{Observations and Data reduction} \label{sec:optimized}

Insight-HXMT \citep{2020SCPMA..6349502Z} launched on 15th June 2017, is China's first X-ray astronomical satellite. HXMT contains three major scientific payloads: High Energy X-ray telescope (HE; 20-250 keV), Medium Energy X-ray telescope (ME; 5–30 keV), and Low Energy X-ray telescope (LE; 1–15 keV), having a time resolution of 25 $\mathrm{\mu} s$, 276 $\rm \mu s$, and 1 ms, respectively. The effective areas of the three payloads are 5000 $\rm cm^2$, 952 $\rm cm^2$, and 384 $\rm cm^2$, respectively.

Insight-HXMT Data Analysis Software (HXMTDAS) v2.04 is used for data reduction. We screen and clean events for all three payloads with the following criteria: (1) the pointing offset angle $<0.04^\circ$; (2) the elevation angle $>10^\circ$; (3) the geomagnetic cut-off rigidity > 8 GeV. In addition, periods when the satellite passes through the South Atlantic Anomaly are excluded, and elevation angle from Earth bright limb $>20^\circ$ was separately applied for LE data. Background light curves are estimated by using hebkgmap, mebkgmap, and lebkgmap tasks (also see details in previous work, \citealt{2021JHEAp..30....1W,2022MNRAS.513.4875C,2022MNRAS.514.2805L}). Tasks helcgen, melcgen, and lelcgen are used to extract X-ray light curves with $\sim$ 0.008 (1/128) sec time bins in the energy bands 30–250 keV, 10–30 keV, and 2–10 keV for HE, ME, and LE, respectively.

\begin{figure}
    \centering
    \includegraphics[width=.5\textwidth]{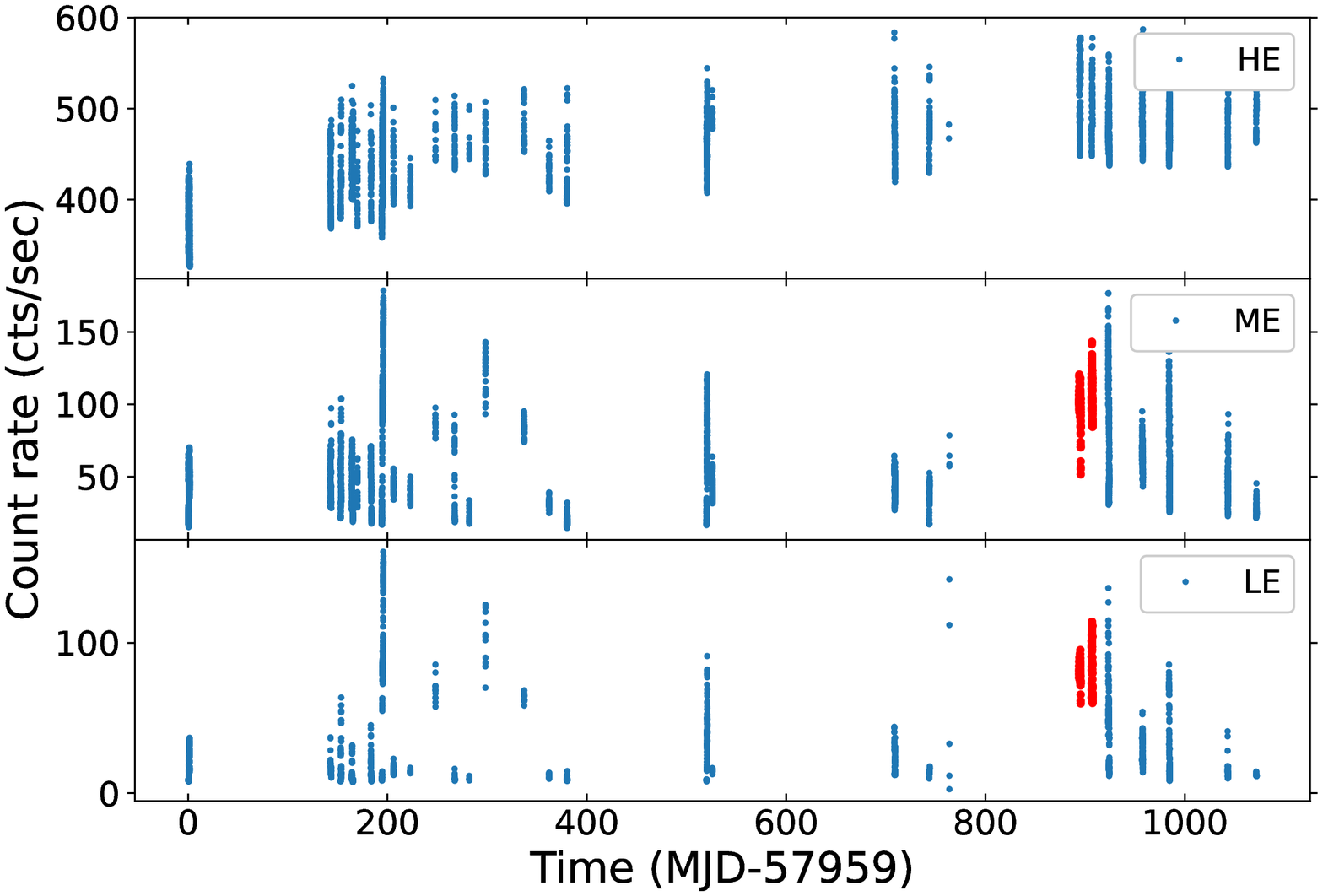}
    \includegraphics[width=.5\textwidth]{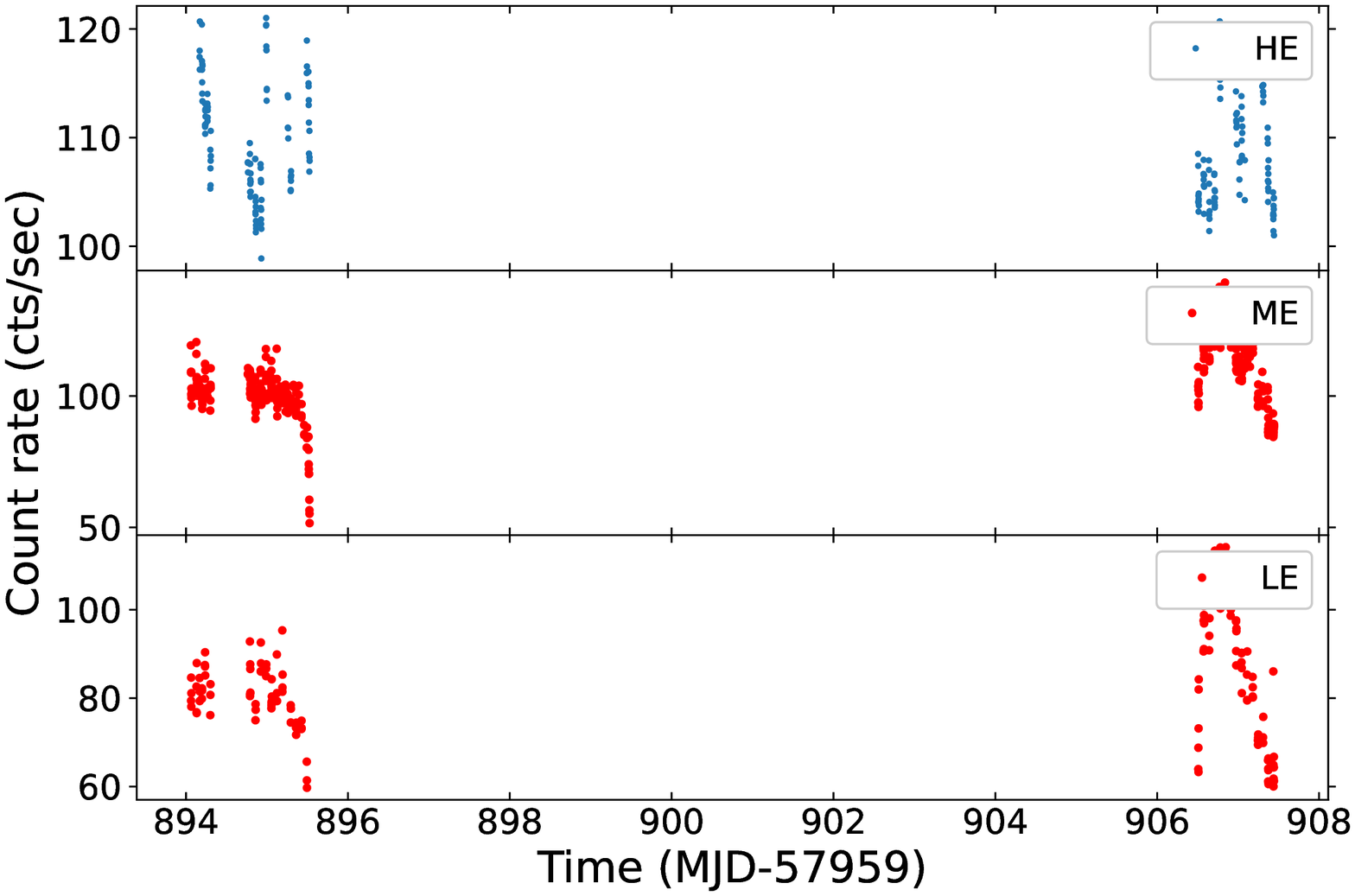}
    \includegraphics[width=.52\textwidth]{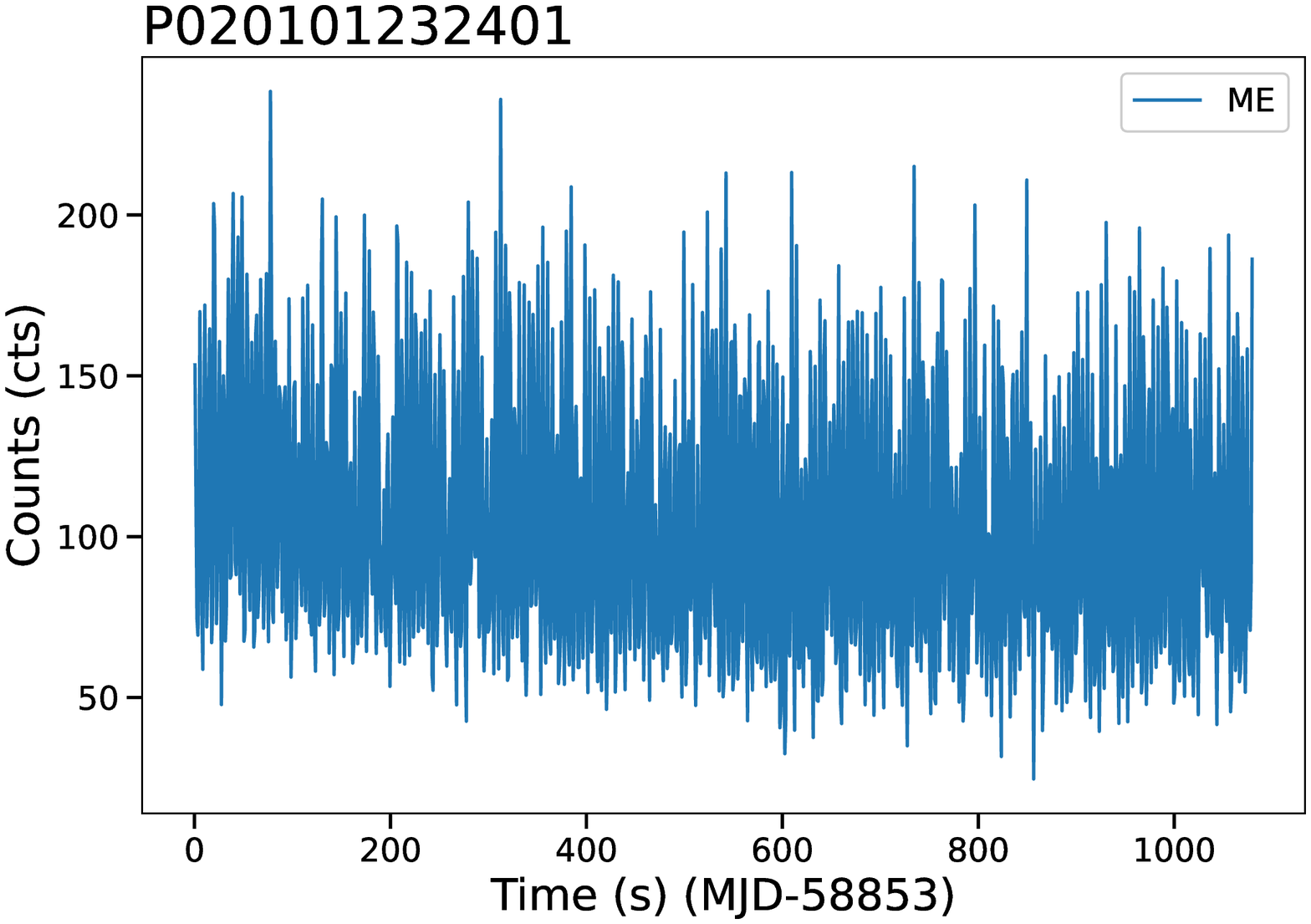}
    \caption{The upper plots are the light curves with a time resolution of 100 s during the Insight-HXMT observations from 2017 to 2020 for LE, ME, and HE. The red points represent the time intervals when the QPO is detected. The middle plots are the zoomed-in light curves of the two red observations. The bottom plots are the count rates for one example observation with the QPO (ExpID P020101232401, ME) with a time resolution of 1 s, where we can clearly see the modulations in the light curve.}
    \label{fig:lc}
\end{figure}

\begin{figure*}
\centering
\includegraphics[width=0.49\textwidth]{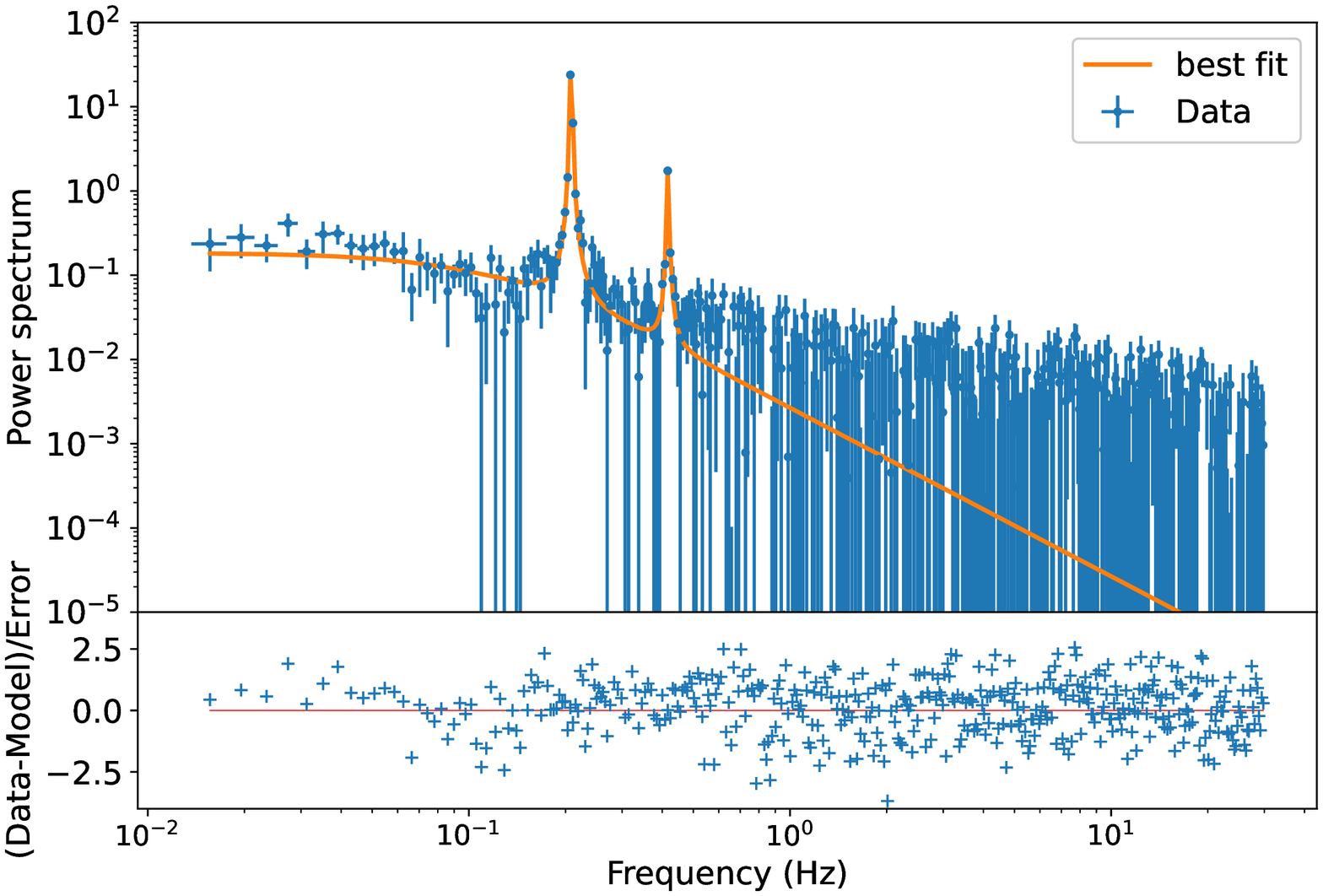}
\includegraphics[width=0.49\textwidth]{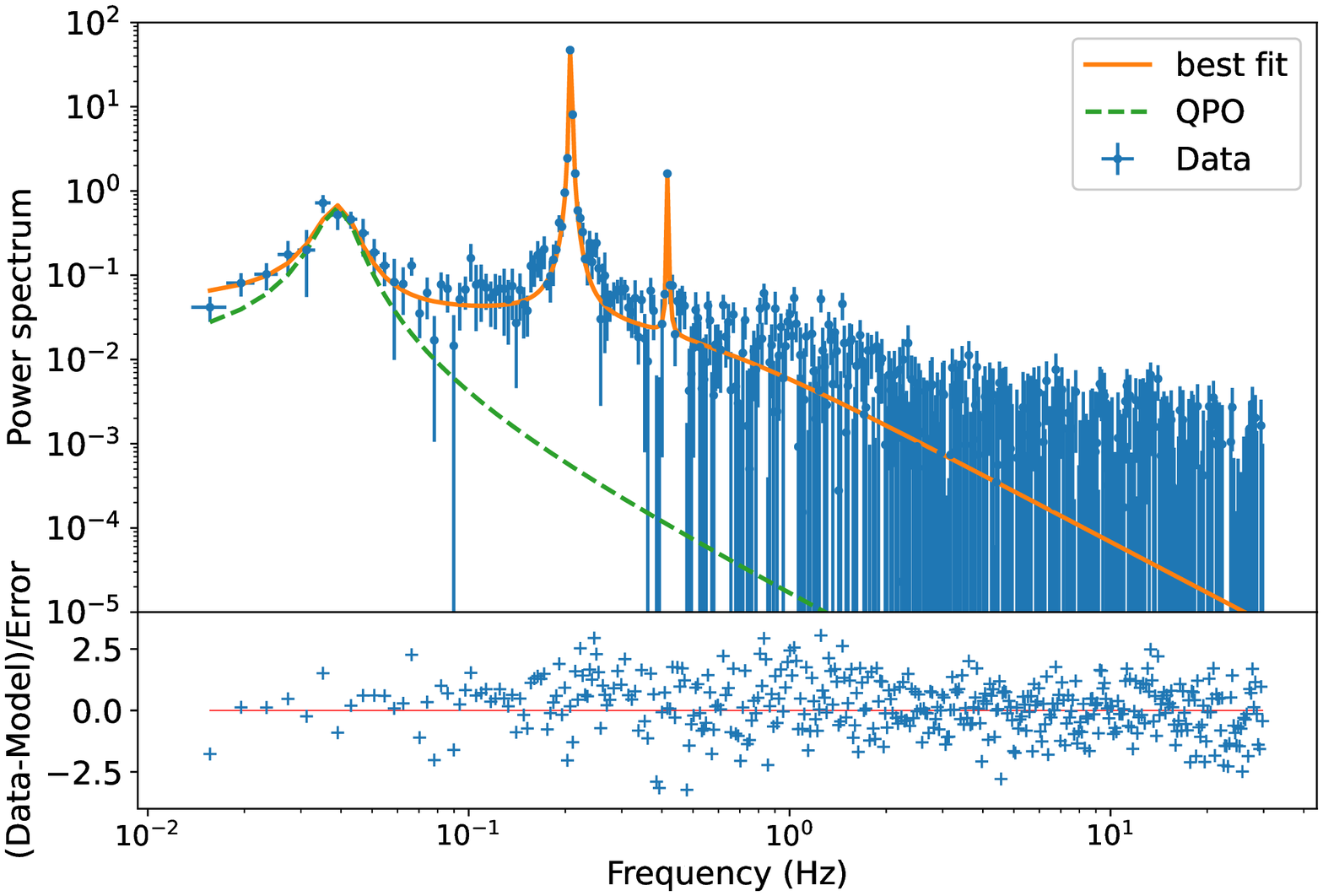}
\caption{The power spectra for ME (10–30 keV) of Cen X-3 were obtained by Insight-HXMT. The left panel shows the example  power spectrum for the observation (P020101232703) without the QPO feature. The right panel (an example based on ExpID P020101232407) shows the model fitting with a QPO, which consists of a broad-band noise and three Lorenztian functions, implying three frequencies: the NS spin frequency $\sim$ 0.2 Hz, the harmonic frequency $\sim 0.4$ Hz, and the QPO frequency $\sim 0.04$ Hz, shown as the solid orange line. The green dashed line represents the QPO feature at $\sim$40 mHz. The fitting residuals are shown in the bottom panels.}
\label{fig:qpo_fit}
\end{figure*}

The light curves of Cen X-3 during 2017-2020 HXMT observations are shown in the top panel of \cref{fig:lc} with a time resolution of 100 s. We visually find obvious QPO features only in two observations labeling in red (ObsIDs: P0201012324; P0201012325) in \cref{fig:lc}, and the zoom-in light curves of these two ObsIDs are also displayed. For the HE data, there is no QPO detected. The detailed information of 2 observations in 2020 in which QPOs are detected in Cen X-3 are summarized in \cref{tab:ObsIDs}. In order to reduce the data file size, Insight-HXMT split each observation into multiple segments (called "exposure") that are identified as the Exposure IDs (ExpID seen in \cref{tab:ObsIDs}). For {one ObsID where QPOs are detected}, some parts of ExpIDs are missing in the table because they have no available data. {The ExpIDs among the two ObsIDs in \cref{tab:ObsIDs}, which have no obvious QPO features,  would be caused by a low Signal-to-Noise ratio or statistic due to short GTIs. For the other ObsIDs, the non-detection of QPOs would have the physical reasons (see the examples in \cref{fig:qpo_fit}).} To perform the timing analysis, the arrival times of photons were corrected from the terrestrial time to the solar system barycentric time using hxbary task in HXMTDAS. 

\begin{table}
    \centering
    \caption{The observation IDs and information of Insight-HXMT when QPOs are detected in Cen X-3. Orbital phases are obtained using the ephemeris of \citealt{2015A&A...577A.130F}. The orbital phases for eclipse ingress and egress at which 99\% of the source flux is occulted are as 0.922 and 0.077, respectively (\citealt{2015A&A...577A.130F}). }
    \label{tab:ObsIDs}
    \begin{tabular}{l|ccccc}
    \hline \hline 
    \multirow{2}{*}{ObsID} & \multirow{2}{*}{Start date} & \multirow{2}{*}{ExpID} & \multirow{2}{*}{Duration [s]} & \multirow{2}{*}{Orbital phase}\\ \\
    \hline 
    P0201012324 & 2020-01-05 & 01 & 9360 & 0.20 \\
                &            & 02 & 9540 & 0.27 \\
                &            & 06 & 8970 & 0.54 \\
                &            & 07 & 6780 & 0.61 \\
                &            & 09 & 8970 & 0.73 \\
                &            & 10 & 8970 & 0.79 \\ \hline
    P0201012325 & 2020-01-17 & 01 & 12300& 0.18 \\
                &            & 02 & 7200 & 0.26 \\
                &            & 03 & 7230 & 0.32 \\
                &            & 05 & 9030 & 0.44 \\
    \hline \hline
    \end{tabular} 
\end{table}

\section{Analysis and results} \label{sec:result}

For the extracted light curves obtained from each payload and exposure, background-subtracted processes were carried out at first. To investigate QPO, we divided the light curves into segments of 256 s and created the power density spectrum (PDS) using the HEASARC tool powspec (XRONOS package) for each exposure. Therefore, PDSs have a frequency resolution of 0.00390625 (1/256) Hz. The PDSs were averaged and Leahy normalized \citep{1983ApJ...272..256L} so that their integral gives the squared RMS fractional variability. The expected white noise level is subtracted, and the uncertainties are evaluated by using the standard deviation of the average. 

\begin{figure}
    \centering
    \includegraphics[width=.5\textwidth]{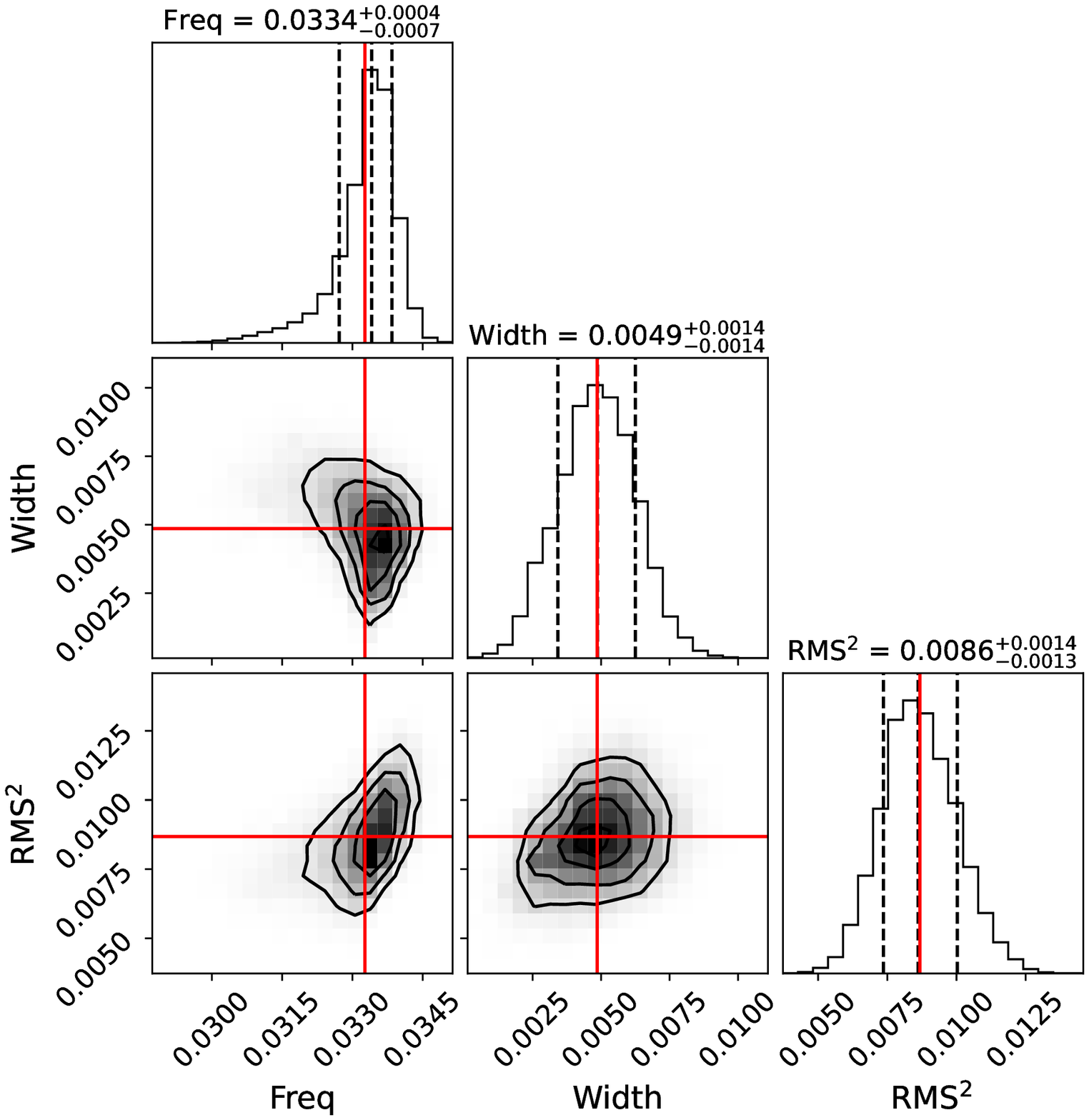}
    \caption{The contour diagram for QPO frequency, width, and RMS$^{2}$ generated by MCMC. There is no apparent parameter degeneration between these parameters.}
    \label{fig:mcmc}
\end{figure}

For the observations without QPO features, the PDSs of both LE (2–10 keV) and ME (10-30 keV) payloads can be characterized by a model generally consisting of a broad-band noise with two spin components. The broad-band noise of PDSs are well described by the Lorentz, and the spin components can be also modelled by the Lorentz function: A(f)=K($\sigma /2 \pi$)/[$(f-f_{c})^2+(\sigma /2)^2$]. \cref{fig:qpo_fit} (left) shows the model fit example of the data without a QPO. For the observations with QPO features (obvious residuals at about 0.04 Hz), another Lorentz function will be added. The acceptable fittings for the LE and ME PDSs are performed based on XSPEC 12.11.1 version \citep{1996ASPC..101...17A}. As seen in \cref{fig:qpo_fit} (right, the example based on the exposure P020101232407), the PDS of ME is comprised of four components: a broadband noise, and two narrow line features (the fundamental and harmonic spin frequencies of the NS), and a broad peak (the QPO feature). Uncertainties of parameters are estimated by the Monte Carlo Markov Chain method with chain steps of 100000. These parameters have no apparent degeneracy seen in \cref{fig:mcmc}. Due to limited statistics (usually GTI $< 1000$ s), the residuals probably have large fluctuations and are more apparent at low frequency. The best-fitting parameters (including QPO frequency, width, and squared rms) are listed in \cref{tab:paras}. {We also use the models including a QPO feature to fit the observation ID (e.g., P020101232703) without reporting the QPO and find that the fitting results (reduced $\chi^2\sim$ 1.19) are consistent with the non-QPO models (reduced $\chi^2\sim$ 1.14). If we perform QPO fittings for this observation assuming the typical QPO frequency (fixed value but ranging from 20 -- 60 mHz) and width (fixed, ranging from 0.005 -- 0.02 Hz), the QPO RMS parameter is float in the fit, and then we give the upper limits of RMS $\sim 7\%$. And the RMS of the QPO for the ObsIDs when it is detected, is always higher than that limit.}

\begin{table*}
    \centering
    \caption{Best-fitting parameters of the LE and ME PDSs, $\nu$ and Width represent the QPO frequency and corresponding width, respectively.}
    \renewcommand\arraystretch{1.5}
    \label{tab:paras}
    \begin{tabular}{l|cccccccc}
    \hline \hline 
    ExpID & Payload &  $\nu$ (Hz) & Width (Hz)& RMS (\%) & reduced $\chi^{2}$ & Orbital phase \\
    \hline 
P020101232401 & LE 	 & $0.033_{-0.001}^{+0.000}$ & $0.005_{-0.001}^{+0.001}$ & 	$9.30_{-0.73}^{+0.69}$  & 1.39 & 0.20\\
..........    & ME 	 & $0.037_{-0.002}^{+0.002}$ & $0.013_{-0.003}^{+0.006}$ & 	$8.76_{-0.95}^{+0.82}$  & 1.36 & 0.20\\
P020101232402 & ME 	 & $0.036_{-0.002}^{+0.002}$ & $0.024_{-0.005}^{+0.007}$ & 	$8.73_{-0.72}^{+0.67}$  & 1.61 & 0.27\\
P020101232406 & LE 	 & $0.040_{-0.002}^{+0.002}$ & $0.015_{-0.005}^{+0.008}$ & 	$11.15_{-1.41}^{+1.31}$ & 1.53 & 0.54\\
.........     & ME 	 & $0.038_{-0.002}^{+0.002}$ & $0.016_{-0.004}^{+0.005}$ & 	$8.31_{-0.64}^{+0.63}$  & 1.25 & 0.54\\
P020101232407 & LE 	 & $0.042_{-0.001}^{+0.001}$ & $0.008_{-0.001}^{+0.001}$ & 	$10.71_{-0.72}^{+0.67}$ & 1.40 & 0.61\\
.........     & ME 	 & $0.039_{-0.001}^{+0.002}$ & $0.011_{-0.002}^{+0.003}$ & 	$10.00_{-0.79}^{+0.71}$ & 1.25 & 0.61\\
P020101232409 & ME 	 & $0.037_{-0.001}^{+0.001}$ & $0.010_{-0.002}^{+0.003}$ & 	$9.47_{-0.85}^{+0.76}$  & 1.21 & 0.73\\
P020101232410 & ME 	 & $0.039_{-0.002}^{+0.002}$ & $0.013_{-0.004}^{+0.004}$ & 	$8.54_{-0.88}^{+0.79}$  & 1.26 & 0.79\\
P020101232501 & LE 	 & $0.037_{-0.001}^{+0.002}$ & $0.010_{-0.002}^{+0.003}$ &  $12.95_{-1.23}^{+1.10}$ & 1.63 & 0.18\\
.........     & ME 	 & $0.039_{-0.001}^{+0.001}$ & $0.013_{-0.003}^{+0.005}$ & 	$9.72_{-0.83}^{+0.73}$  & 1.57 & 0.18\\
P020101232502 & LE 	 & $0.038_{-0.002}^{+0.002}$ & $0.016_{-0.004}^{+0.007}$ & 	$9.54_{-0.75}^{+0.73}$  & 1.39 & 0.26\\
.........     & ME 	 & $0.037_{-0.001}^{+0.001}$ & $0.009_{-0.002}^{+0.004}$ & 	$8.14_{-0.81}^{+0.77}$  & 1.48 & 0.26\\
P020101232503 & LE 	 & $0.034_{-0.001}^{+0.001}$ & $0.009_{-0.002}^{+0.002}$ & 	$9.77_{-0.70}^{+0.65}$  & 1.27 & 0.32\\
.........     & ME 	 & $0.034_{-0.001}^{+0.001}$ & $0.009_{-0.002}^{+0.003}$ & 	$7.98_{-0.69}^{+0.61}$  & 1.57 & 0.32\\
P020101232505 & ME 	 & $0.038_{-0.002}^{+0.002}$ & $0.013_{-0.003}^{+0.004}$ & 	$7.90_{-0.89}^{+0.80}$  & 1.43 & 0.44\\
    \hline \hline
    \end{tabular} 
\end{table*}

\begin{figure}
    \centering
    \includegraphics[width=.5\textwidth]{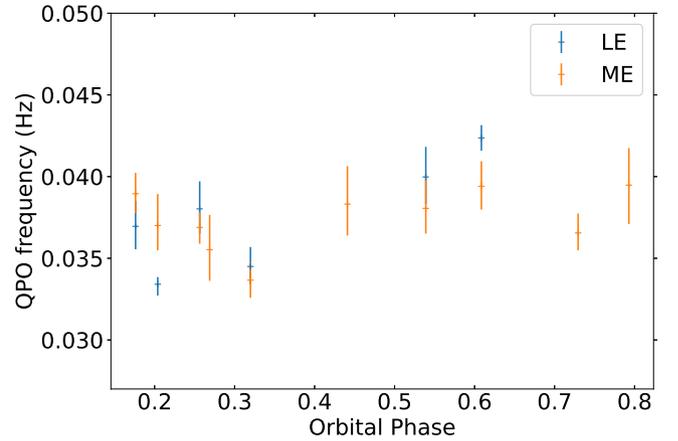}
    \caption{QPO frequency versus orbital phases. Blue and orange points are shown as LE and ME results, respectively.}
    \label{fig:QPOfreq}
\end{figure}

\begin{figure}
    \centering
    \includegraphics[width=.5\textwidth]{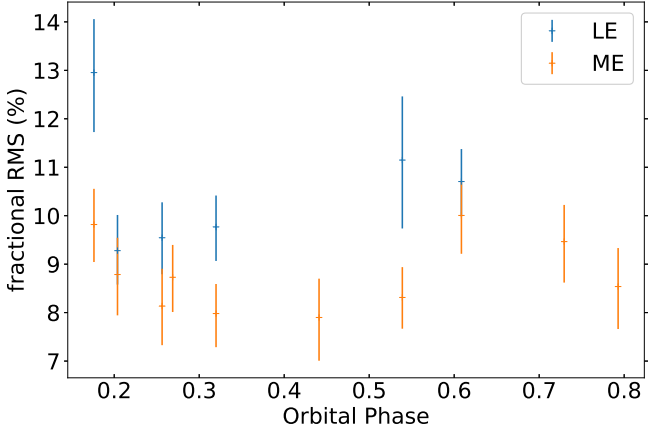}
    \caption{QPO rms amplitude versus orbital phases. Blue and orange points are shown as LE and ME results, respectively.}
    \label{fig:QPOrms}
\end{figure}


\begin{table*}
\centering
\caption{The best-fitting spectral parameters of Cen X-3 in the hard X-ray bands from 2–100 keV for the ExpIDs with QPO detection. The best model is described with the power-law multiplied by high energy cut-off (highec, XSPEC), a cyclotron line (gabs, XSPEC), an Fe K$\alpha$ line fixed at 6.67 keV, and the absorbtion column density (TBabs, XSPEC). $N_{H}$ and $\Gamma$ represent the neutral hydrogen column density and photon index, respectively. $E_{\rm cycl}$, $W_{\rm cycl}$ and $D_{\rm cycl}$ are the centroid energy, the width, and the depth of the cyclotron line. The flux is given in units of erg cm$^{-2}$ s$^{-1}$, and the absorption column density in units of 10$^{22}$ atoms cm$^{-2}$. The other parameters $E_{\rm c}$ (cutoff energy), $\ E_{\rm f}$ (folding energy), $\ E_{\rm cycl}$,  $\ W_{\rm cycl},$ and $\sigma_{\rm Fe}$ are in units of keV. Uncertainties are given at the 68\% confidence.}
\label{tab:specfit}
\renewcommand\arraystretch{1.8}
\setlength{\tabcolsep}{0.45mm}{
\begin{tabular}{l|ccccccccccccccc}
\hline
ExpID & $N_{H}$& $\Gamma$ & $E_{\rm c}$ & $E_{\rm f}$ & $E_{\rm cycl}$ & $W_{\rm cycl}$ & $D_{\rm cycl}$  & $\sigma_{\rm Fe}$ & Log Flux & $\chi^2$/dof\\ 
\hline 
P020101232401 &  	$1.47_{-0.09}^{+0.09}$ & $1.18_{-0.02}^{+0.02}$ &  $11.19_{-0.28}^{+0.23}$ & $10.33_{-0.50}^{+0.70}$ & $31.20_{-0.92}^{+0.57}$ & $4.72_{-0.72}^{+0.75}$ & $10.99_{-2.24}^{+2.18}$ & $0.22_{-0.04}^{+0.04}$ &  	$-8.063_{-0.005}^{+0.005}$ & 1288/1310 \\ \hline   
P020101232402 &  	$1.48_{-0.08}^{+0.08}$ & $1.18_{-0.02}^{+0.02}$ &  $11.24_{-0.24}^{+0.22}$ & $11.39_{-0.40}^{+0.40}$ & $29.41_{-0.79}^{+1.04}$ & $4.93_{-0.48}^{+0.52}$ & $8.38_{-1.30}^{+1.52}$ & $0.42_{-0.06}^{+0.06}$ &  	$-8.042_{-0.002}^{+0.003}$ & 1267/1310 \\ \hline
P020101232406 &  	$1.17_{-0.08}^{+0.08}$ & $1.12_{-0.02}^{+0.02}$ &  $11.19_{-0.22}^{+0.20}$ & $11.08_{-0.38}^{+0.38}$ & $28.18_{-0.73}^{+0.88}$ & $5.00_{-0.45}^{+0.55}$ & $7.27_{-1.09}^{+1.22}$ & $0.33_{-0.04}^{+0.05}$ &  	$-8.035_{-0.002}^{+0.003}$ & 1251/1310 \\ \hline
P020101232407 &  	$1.19_{-0.11}^{+0.11}$ & $1.15_{-0.03}^{+0.03}$ &  $10.36_{-1.24}^{+0.47}$ & $12.63_{-0.53}^{+0.52}$ & $27.05_{-0.52}^{+0.64}$ & $4.45_{-0.30}^{+0.35}$ & $8.53_{-0.87}^{+0.90}$ & $0.26_{-0.05}^{+0.06}$ &  	$-8.017_{-0.003}^{+0.003}$ & 1237/1310 \\ \hline
P020101232409 &  	$0.92_{-0.12}^{+0.13}$ & $0.98_{-0.04}^{+0.04}$ &  $9.20_{-0.36}^{+0.36}$ & $11.29_{-0.44}^{+0.51}$ & $29.63_{-1.14}^{+1.41}$ & $5.79_{-0.63}^{+0.66}$ & $11.76_{-2.40}^{+2.99}$ & $0.27_{-0.07}^{+0.10}$ &  	$-8.037_{-0.004}^{+0.004}$ & 1178/1310 \\ \hline
P020101232410 &  	$1.58_{-0.13}^{+0.12}$ & $1.25_{-0.03}^{+0.03}$ &  $11.85_{-0.29}^{+0.26}$ & $10.50_{-0.58}^{+0.64}$ & $29.76_{-1.79}^{+1.54}$ & $4.91_{-1.12}^{+1.07}$ & $5.18_{-2.02}^{+2.26}$ & $0.52_{-0.09}^{+0.10}$ &  	$-8.100_{-0.004}^{+0.004}$ & 1226/1310 \\ \hline
P020101232501 &  	$1.99_{-0.06}^{+0.06}$ & $1.17_{-0.01}^{+0.01}$ &  $11.39_{-0.18}^{+0.17}$ & $11.02_{-0.28}^{+0.29}$ & $29.79_{-0.73}^{+0.93}$ & $4.56_{-0.39}^{+0.46}$ & $8.67_{-1.11}^{+1.55}$ & $0.40_{-0.06}^{+0.08}$ &  	$-7.998_{-0.002}^{+0.002}$ & 1290/1310 \\ \hline
P020101232502 &  	$1.45_{-0.08}^{+0.07}$ & $1.09_{-0.02}^{+0.02}$ &  $8.25_{-0.21}^{+0.25}$ & $11.61_{-0.25}^{+0.26}$ & $29.50_{-0.65}^{+0.81}$ & $4.76_{-0.36}^{+0.42}$ & $9.52_{-1.11}^{+1.44}$ & $0.23_{-0.03}^{+0.03}$ &	$-7.932_{-0.002}^{+0.002}$ & 1324/1310 \\ \hline
P020101232503 &  	$1.65_{-0.08}^{+0.08}$ & $1.17_{-0.02}^{+0.02}$ &  $9.21_{-0.50}^{+0.84}$ & $16.56_{-0.83}^{+0.89}$ & $31.18_{-0.91}^{+0.61}$ & $6.91_{-0.46}^{+0.37}$ & $20.16_{-2.64}^{+2.38}$ & $0.30_{-0.03}^{+0.03}$ &  	$-7.904_{-0.006}^{+0.006}$ & 1136/1310 \\ \hline
P020101232505 &  	$1.43_{-0.09}^{+0.09}$ & $1.18_{-0.02}^{+0.02}$ &  $11.76_{-0.27}^{+0.26}$ & $12.16_{-0.68}^{+0.71}$ & $29.41_{-1.12}^{+1.35}$ & $5.01_{-0.74}^{+0.72}$ & $8.18_{-1.91}^{+2.29}$ & $0.29_{-0.05}^{+0.05}$ &  $-8.019_{-0.004}^{+0.004}$ & 1337/1310 \\ \hline
	
\end{tabular} 
}
\end{table*}

\begin{figure}
    \centering
    \includegraphics[width=.5\textwidth]{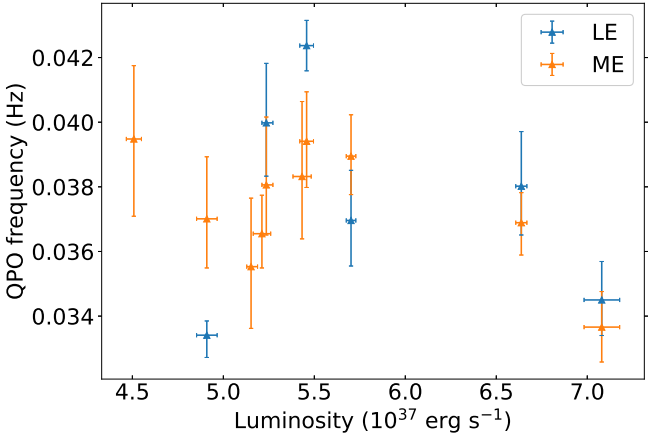}
    \caption{Oscillation frequency versus 2–100 keV X-ray luminosity of Cen X-3. The LE observation is marked by a blue triangle and ME with an orange triangle.}
    \label{fig:lumi}
\end{figure}

\begin{figure}
    \centering
    \includegraphics[width=.5\textwidth]{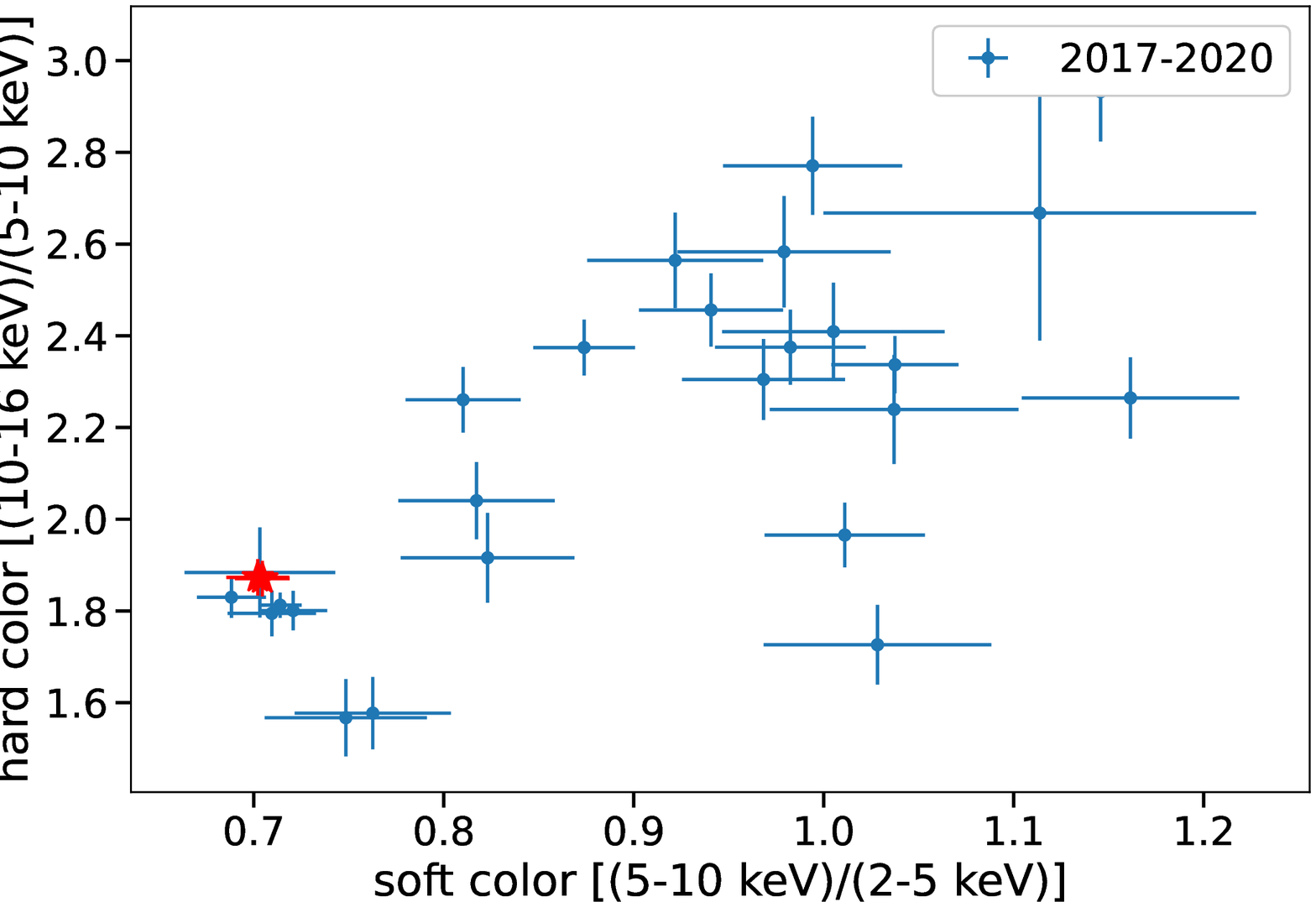}
    \caption{Color-color diagram of Cen X-3 using all available HXMT observations. Red stars mark the position of the observations where mHz QPOs are detected.}
    \label{fig:ccd}
\end{figure}

We first study the evolution of QPO properties with the orbital phases. The average value of QPO frequency is $\sim$39 mHz with large variations. As can be seen from \cref{fig:QPOfreq}, the QPO frequency has the value of $\sim$33 to 39 mHz at the phase of 0.1-0.4, then increases to a higher value range ($\sim$37 --43 mHz) at the orbital phases of 0.4-0.8. The width (full width at half maximum, FWHM) of QPO has an average value of $\sim$ 0.01 Hz, still varies with the orbital phase. Here we define the quality factor $Q=\nu/FWHM$ to describe the QPO characteristics, then we find that $Q\sim 2-7$ for the phase of 0.1 --0.4, while $Q\sim 2-5$ around phase 0.4 -- 0.8. The fractional rms amplitude of QPOs has no apparent evolution with orbital phases and the value varies from $\sim (7- 10)\%$ for ME, and $\sim (9-13)\%$ for LE.

To estimate the flux of the neutron star, we also study the X-ray spectra of Cen X-3 based on LE, ME and HE detectors covering a wide energy band from 2 -- 100 keV. To model the continuum spectra, we tried many models, such as a simple power-law with
high energy exponential roll-off (cutoffpl, XSPEC), Fermi-Dirac cut-off model, or the Negative Positive Exponential (NPEX) model, and find that a power-law with the high energy cut-off model can describe it well. For the cyclotron absorption line component, we used a gaussian absorption line model (gabs, XSPEC). Therefore, the best models consist of a continuum spectrum ($pow*highecut$ models in XSPEC \citealt{1996ASPC..101...17A}), an iron emission line fixed at 6.67 keV, a cyclotron absorption line $gabs$ at $\sim$ 30 keV if present, and the absorption column density ($TBabs$, XSPEC). 
The best fitting parameters of all the observations where QPOs are detected are presented in \cref{tab:specfit}. {A full spectral analysis will be performed and spectral variation properties would be presented in a forthcoming paper.} The luminosity is computed from the X-ray flux using the source distance of $\sim$6.9 kpc \citep{2021MNRAS.507.3899V}. The relation of QPO frequency with luminosity can be seen in \cref{fig:lumi}. There is no apparent dependence of oscillation frequency on X-ray luminosity ($L_{2-100}$ in a narrow range of $\sim (4.5-7.1) \times$ 10$^{37}$ erg \ s$^{-1}$). The luminosity may not represent the true mass accretion rate of Cen X-3 due to a varying degree of obscuration by an aperiodically precessing warped accretion disk \citep{2008MNRAS.387..439R}.

We also study the spectral states of Cen X-3 when the QPOs are detected as shown in \cref{fig:ccd}, where red stars represent the observations that QPOs are observed. Compared with the other Cen X-3 observations of Insight-HXMT from 2017 -- 2020, the X-ray colors reveal that Cen X-3 is in a soft spectral state when it exhibits the QPOs. It may indicate that the generation of QPO is more likely to be related to the soft photon components such as the disc. It should be pointed out that the QPO detections would depend on the count rates, for the hard state, we may not find the QPOs due to the low rates for LE and ME.

\begin{figure*}
    \centering
    \includegraphics[width=.49\textwidth]{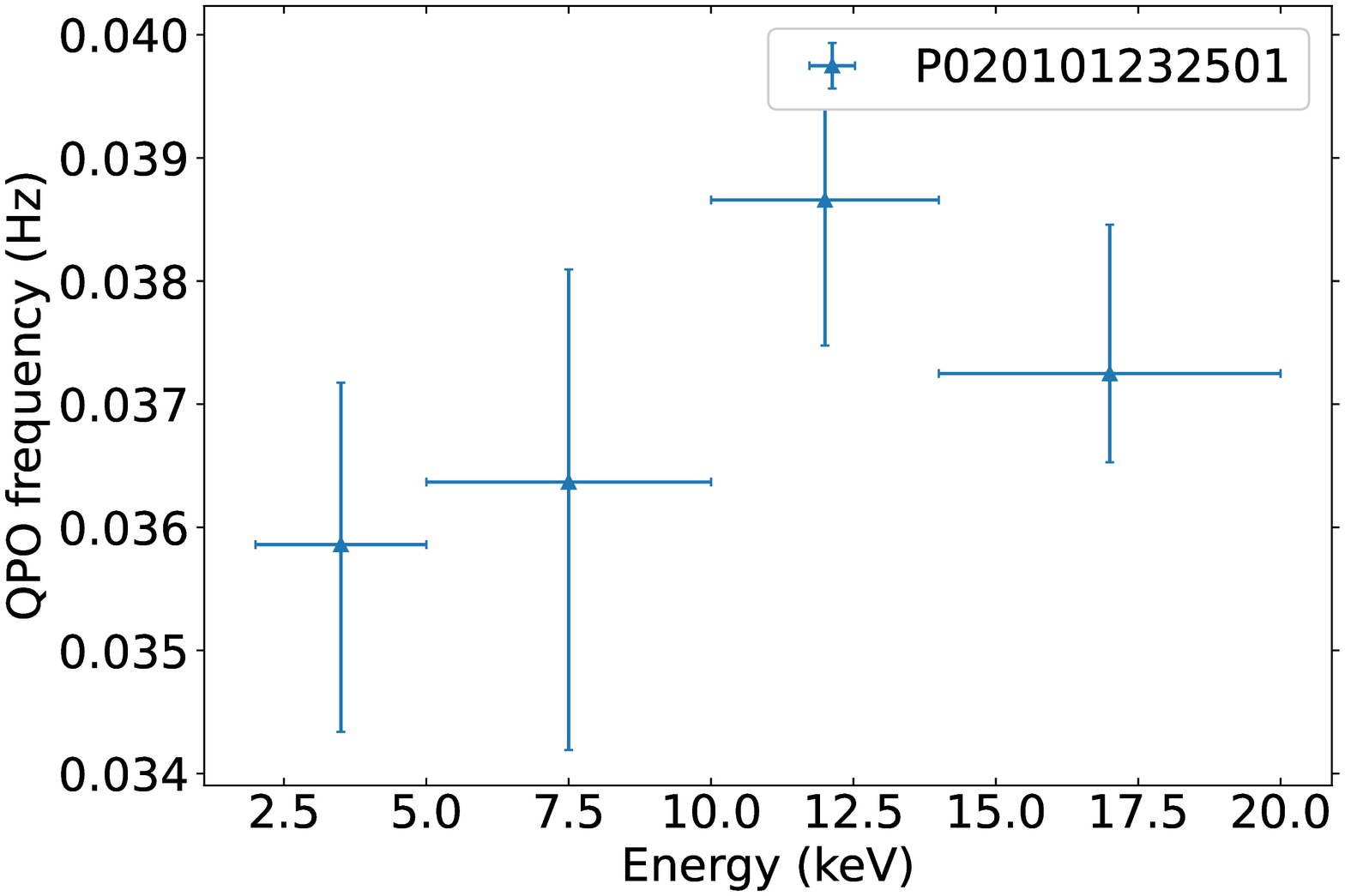}
    \includegraphics[width=.49\textwidth]{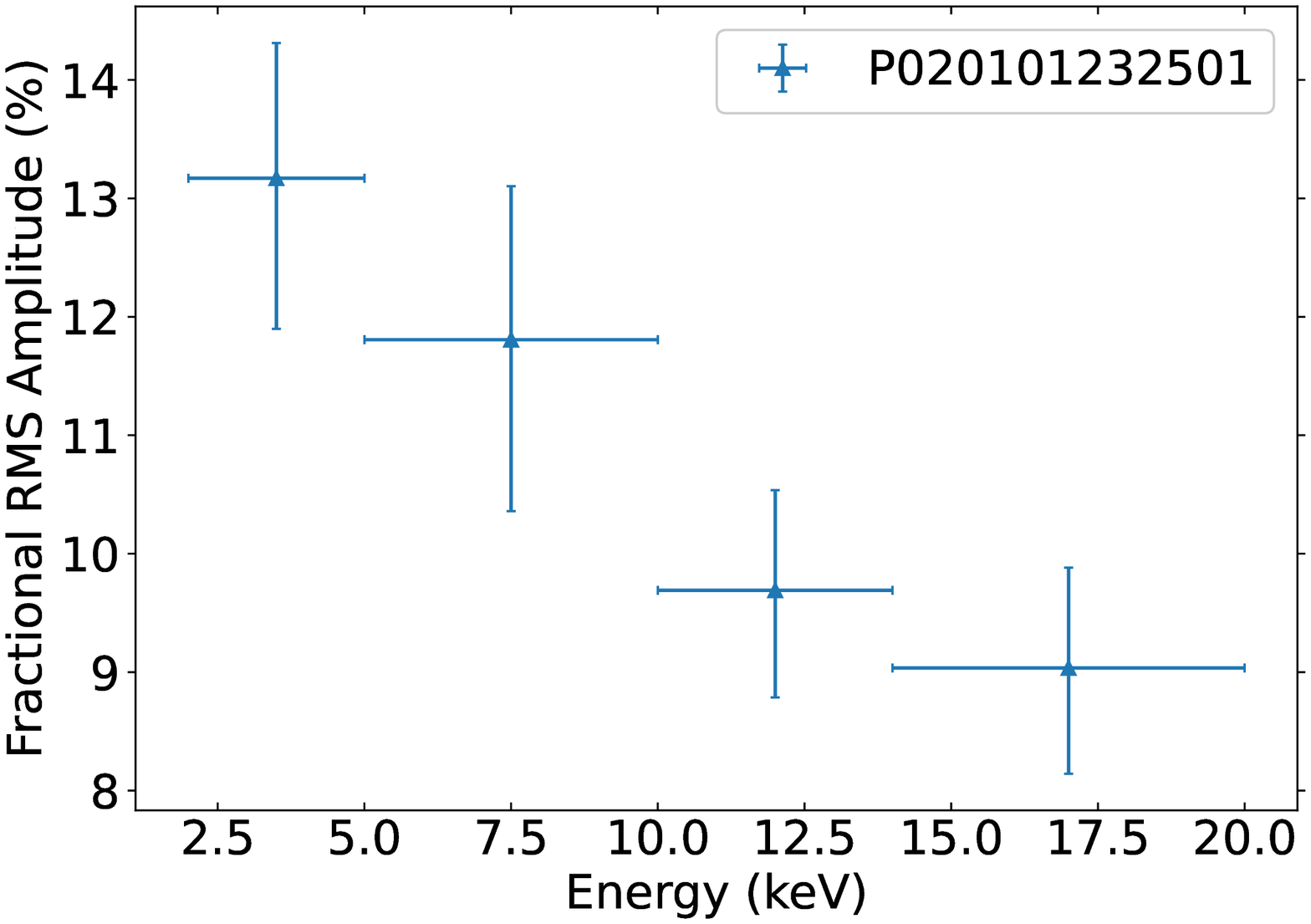}
    \includegraphics[width=.49\textwidth]{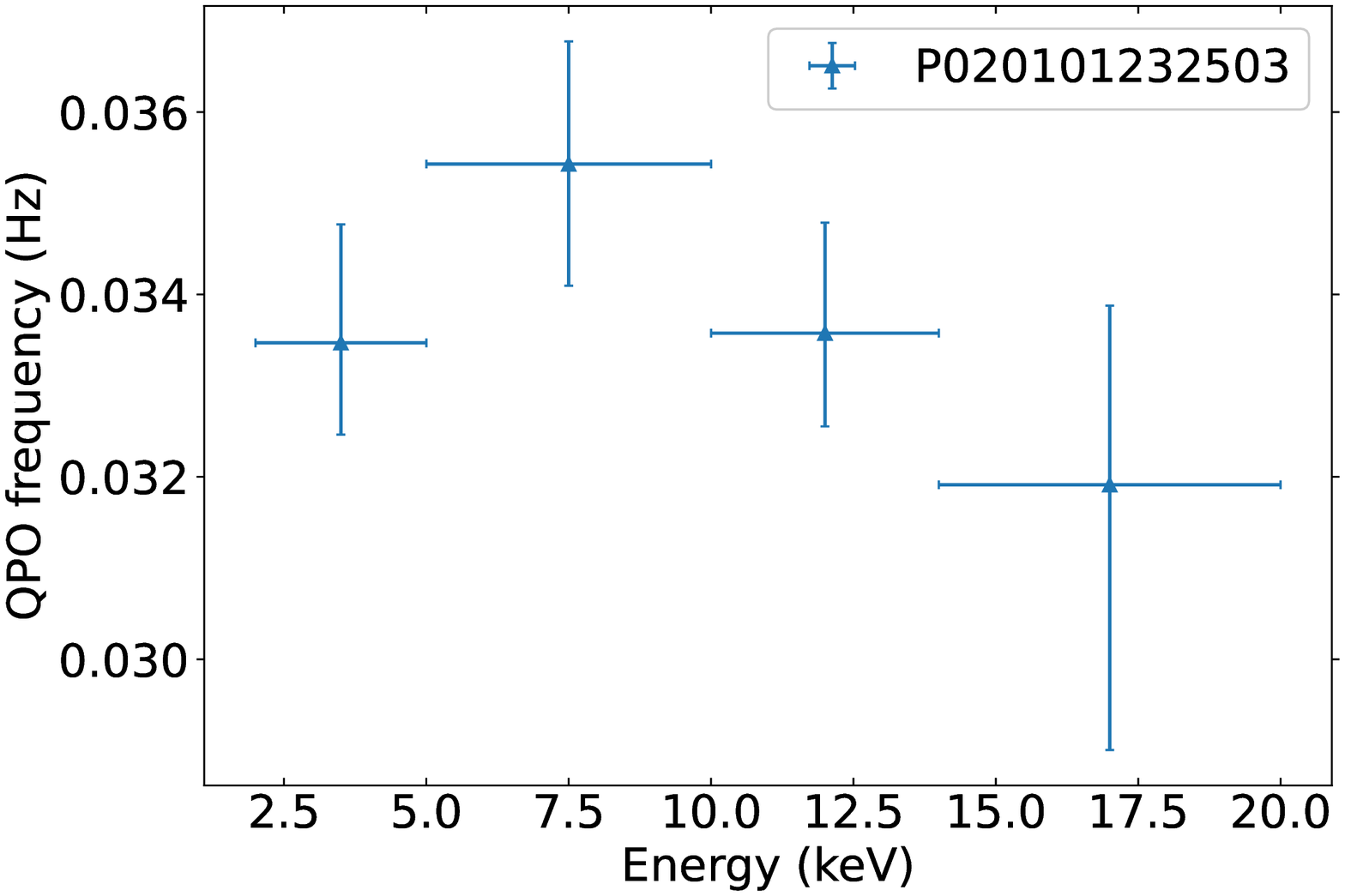}
    \includegraphics[width=.49\textwidth]{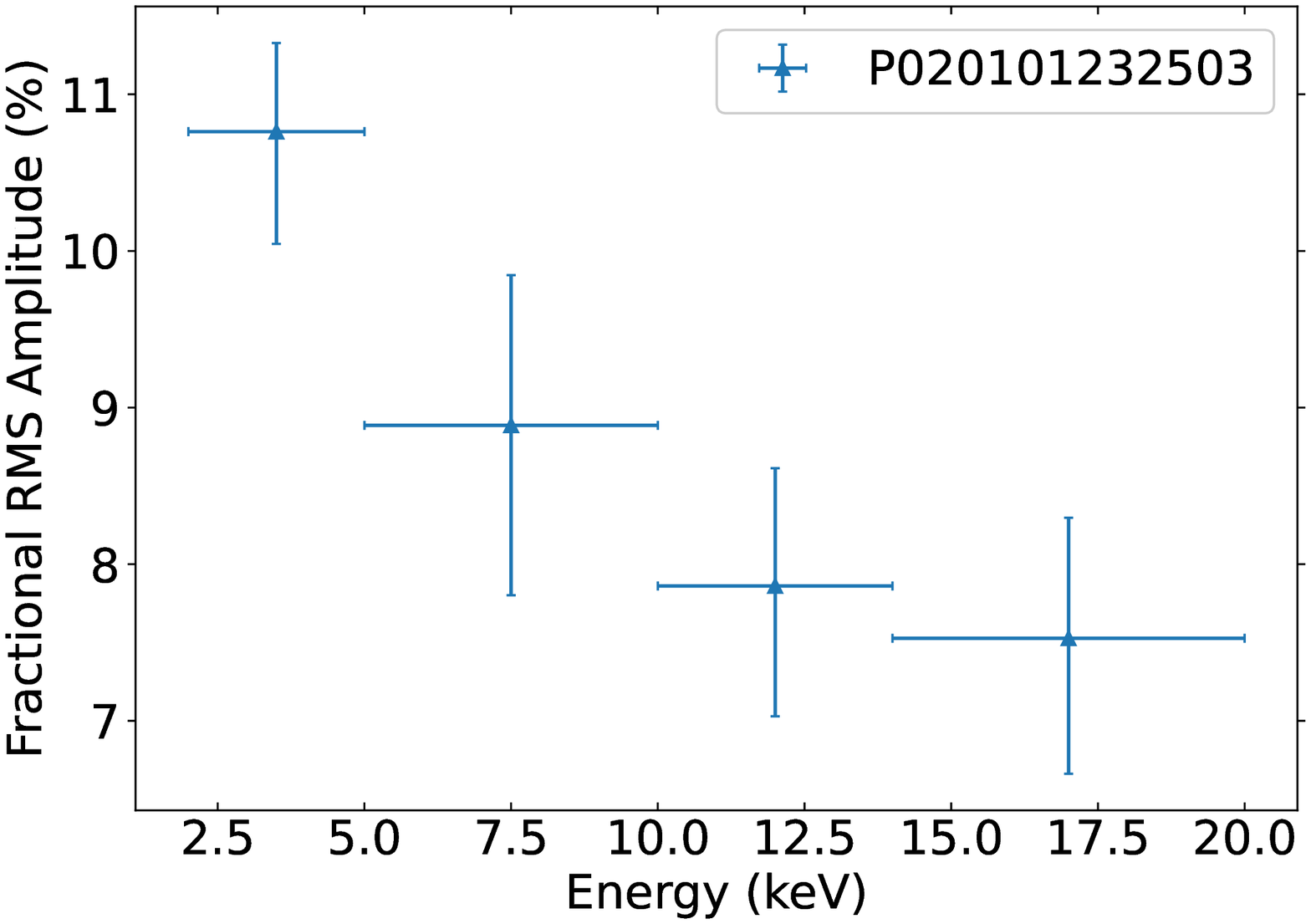}
    \caption{The variations of QPO frequency and rms fractional amplitude are shown here as a function of energy. The $\sim$40 mHz QPO is not detected beyond 20 keV.}
    \label{fig:QPOenergy}
\end{figure*}

For the segments where QPOs are detected in both LE and ME payloads, we find that the fractional rms of QPOs for LE are larger than that for ME. Therefore, we further investigate the energy dependence of the QPO frequency and rms with more sub-energy bands. We generate the PDSs in the energy bands of 2-5 keV and 5-10 keV for LE data and 10-14 keV, 14-20 keV, and 20-30 keV for ME data in two ExpIDs (P020101232501,P020101232503). Then an acceptable fit for each PDS is carried out. We calculate the frequency and fractional rms of the QPO, as seen in \cref{fig:QPOenergy}. It is noted that there is no QPO feature detected above 20 keV.

As shown in \cref{fig:QPOenergy}, QPO frequency has a large variation with the increasing energies for both two observational segments. As the energy increases, the rms amplitude decreases. It varies from 13\% to 9\% for the exposure of P020101232501 and ranges from 11\% to 7\% for P020101232503. It is interesting that QPO rms amplitude in Cen X-3 has an anti-correlation with energy. Our results are different from the previous studies \citep{2008ApJ...685.1109R} which suggest that the rms variation of the 40 mHz QPO is not dependent on the X-ray energy.

\begin{figure}
    \centering
    \includegraphics[width=.5\textwidth]{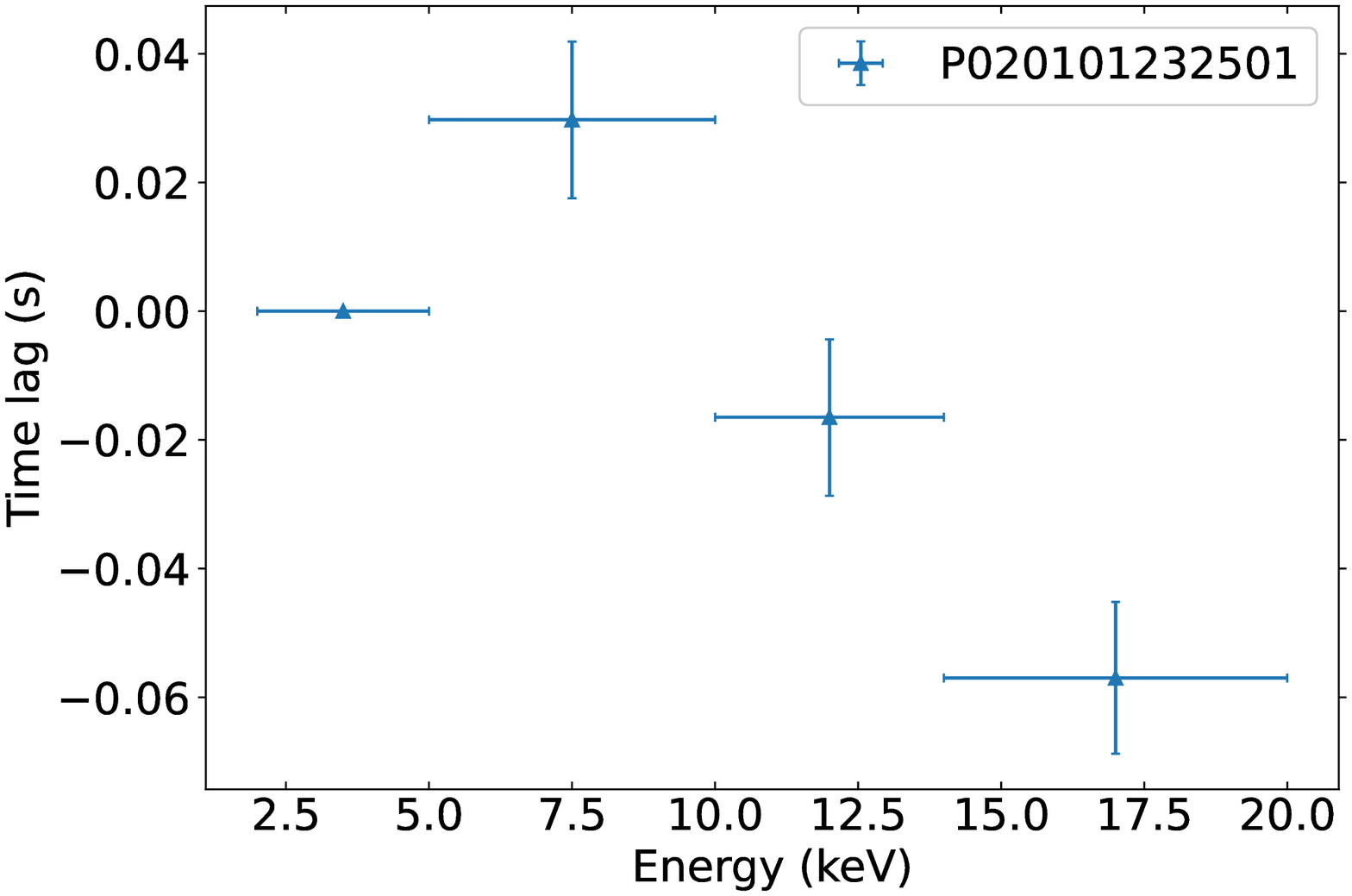}
    \includegraphics[width=.5\textwidth]{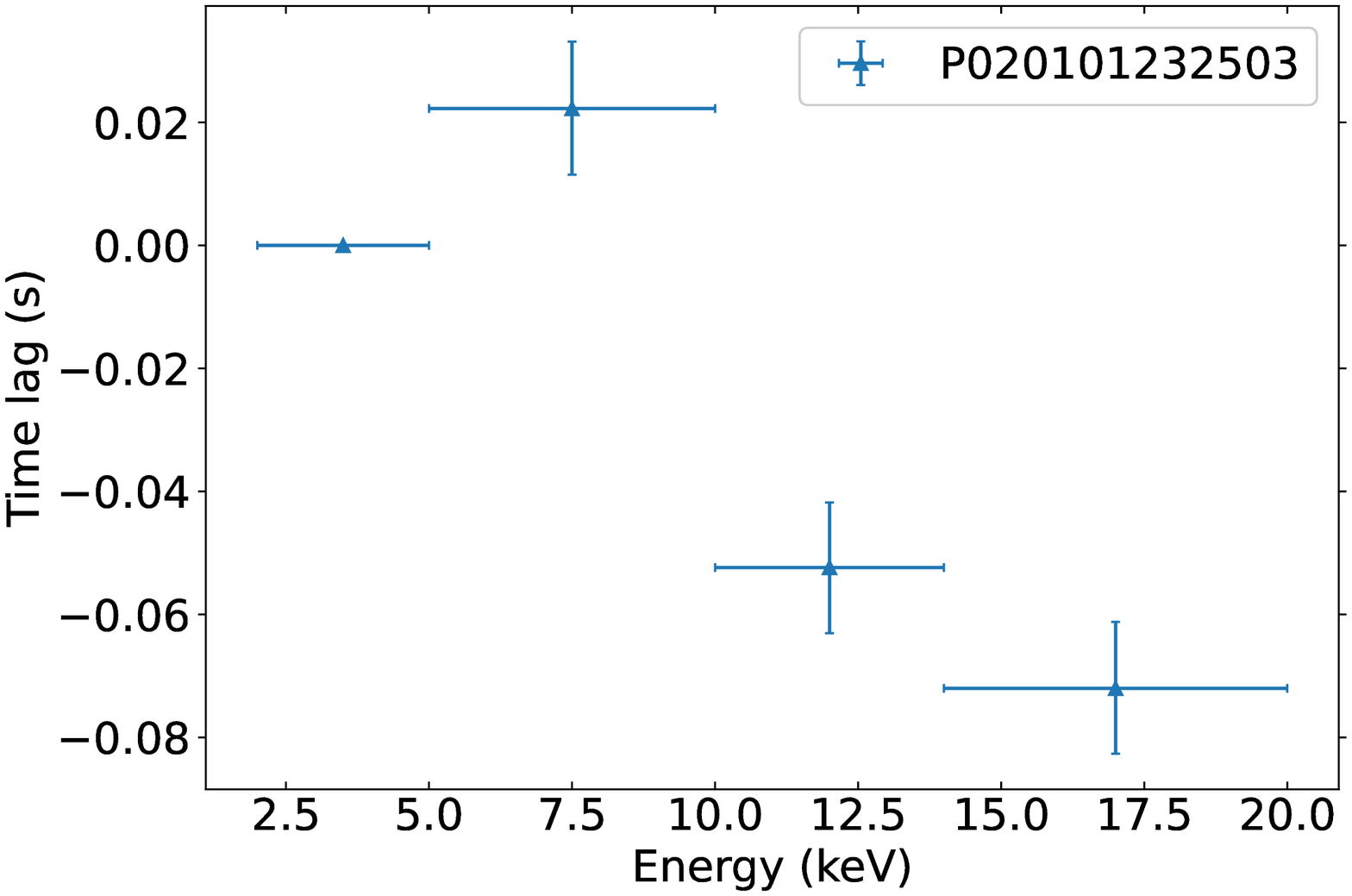}
    \caption{Time delay as a function of photon energy for P020101232501 and P020101232503, respectively.}
    \label{fig:lagenergy}
\end{figure}

\cref{fig:lagenergy} shows the corresponding energy-dependent time lags in 2-5 keV, 5-10 keV, 10-14 keV, and 14-20 keV for the exposure P020101232501 and P020101232503 as an example, respectively, where the reference energy band is taken to be 2–5 keV. The time-lag is calculated using the cross correlation algorithm (crosscor in HEASoft) between a light curve and the reference. The accuracy of time-lag can be obtained by fitting the time-lag versus cross-cor with a gaussian function. We find that a positive time tag of $\sim 20$ ms is presented at 5-10 keV (a hard lag), and then the time lag decreases and becomes negative ($\sim - (20-70)$ ms, soft lag) in the higher energies. These time-lag behaviors seen in one observation ID is similar, while the time-lag behaviors have a little difference for two different obsIDs. 

\section{Discussion} \label{sec:discussion}


The 40 mHz QPO features have been observed from the HMXB pulsar Cen X-3 in two observations made with Insight-HXMT in 2020. Besides, the two observations cross over approximately one orbital period, the QPO frequency evolved with orbital phases from $\sim 33$ mHz around the orbital phase 0.1 to $\sim 40$ mHz near the phases of $\sim 0.8$. {In addition, the QPOs can not be detected in the other Insight-HXMT observations yet, then the appearance and disappearance of mHz QPOs could be connected to the physical origins.} The origin of QPOs in the neutron star X-ray binary is still uncertain, here we will discuss different theoretical models including KFM, BFM, the thermal disk instabilities, and magnetic disk precession model, which however cannot account for the mHz QPO feature in Cen X-3. We also introduce a possible scenario of the instability near the inner edge of the truncated disc (close to the corotation radius) to explain the mHz QPO.

Two popular models including the Keplerian frequency model \citep{1987ApJ...316..411V} and the Beat frequency model \citep{1985Natur.316..239A} have been proposed to explain the presence of QPOs in X-ray pulsars. In the Keplerian frequency model (KFM), the QPO frequency is modulated by inhomogeneities in the inner accretion disc at the Keplerian frequency and therefore, is the same as that of the Keplerian orbital frequency, such as A 0535+262 \citep{1996ApJ...459..288F}. In the beat frequency model (BFM), this oscillation producing at the beat frequency between the Keplerian frequency at the Alfv$\acute{\mathrm{e}}$n radius and the neutron star spin frequency, is equal to the difference between the Keplerian frequency and the spin frequency. As an example, EXO 2030+375 \citep{1989ApJ...346..906A} was found to be in good agreement with the beat-frequency model where the Keplerian frequency is modulated by the rotating magnetic field of the NS.  

Here, we reported a low frequency quasi-periodic oscillation of $\sim$40 mHz and a spin frequency of $\sim$210 mHz in Cen X-3. The KFM can not be applicable because the QPO frequency is below the neutron star spin frequency and accreted matter may be expelled from the system due to centrifugal inhibition. Therefore, we decided to check the BFM. The Alfv$\acute{\mathrm{e}}$n radius is given by \citep{1973ApJ...184..271L,2012A&A...544A.123B}
\begin{equation}
\begin{aligned}
R_{\mathrm{A}}=& 2.73 \times 10^{7} \times \left(\frac{\Lambda}{0.1}\right) M_{1.4}^{1/7}R_{6}^{10/7} B_{12}^{4/7} {L_{37}^{-2/7}} \mathrm{~cm} ,
\end{aligned}
\end{equation}
where $L_{37}$ is the X-ray luminosity in the units of 10$^{37}$ erg $\cdot$ s$^{-1}$ ($L_{2-100}$ ranging from $\sim 4.5-7.1 \times 10^{37}$ erg $\cdot$ s$^{-1}$ in our paper seen in \cref{fig:lumi}) and $B_{12}$ is magnetic field strength in units of 10$^{12}$ G for a neutron star. We note that the mass and radius of NS are typically $\sim$ 10$^6$ cm and 1.4 M$_{\odot}$, respectively. $\Lambda$ = 1 for spherical accretion and $\Lambda$ < 1 for disk accretion. Based on Eq. (19)
from \cite{2012A&A...544A.123B}, $\Lambda$ can be approximated as 0.1 in the disk case. The Alfven radius $R_{\mathrm{A}}$ in the disk and spherical cases are about $3 \times 10^{7}$ cm and $3 \times 10^{8}$ cm for the typical X-ray luminosity, respectively. Keplerian frequency at the magnetospheric radius can be estimated by
\begin{equation}
v_{\mathrm{k}}=\left(\frac{G M} {4 \pi^{2} R_{\mathrm{A}}^{3}}\right)^{\frac{1}{2}} ,
\end{equation}
it is located at $\sim$ 15 Hz in disk case or $\sim$ 500 mHz in spherical case. The 15 Hz Keplerian frequency is about two orders of magnitude larger than the spin frequency ($\sim$ 0.21 Hz). Taking the Keplerian frequency of 500 mHz, the QPO frequency would be expected to be $v_{\mathrm{qpo}}$ = $v_{\mathrm{k}}$ - $v_{\mathrm{spin}}$ $\approx$ 300 mHz. It is rather larger than the oscillation frequency in measurements. It is also evident from \cref{fig:lumi} that the QPO frequency of Cen X-3 does not show the luminosity dependence as expected in the BFM.

The other possible explanation for the oscillations is caused by thermal disk instabilities occurring in the inner accretion region. The thermal instability can be used to explain QPOs involving the viscous timescales of a few minutes to hours. Viscous timescales can be given as in \cite{2002apa..book.....F}
\begin{equation}
t_{\mathrm{visc}}=991.1 \times L_{37}^{-23 / 35}\left(\frac{\alpha}{0.1}\right)^{-109 / 230} \mathrm{~s} ,
\end{equation}
where $\alpha$ is the disk viscosity parameter \citep{1973A&A....24..337S}. But this model can only explained the QPO frequency of the order $\sim$ 1 mHz.

Otherwise, we consider another warping and precession model in a viscous accretion disk proposed by \cite{2002ApJ...565.1134S}. They conclude that warping and precession modes are caused by magnetic torques and can give rise to variabilities or quasi-periodic oscillations. The corresponding QPO precessional timescales \citep{2019ApJ...872...33R} is given by 
\begin{equation}
\tau_{\text {prec }}=776 \times (\frac{\alpha}{0.1})^{-0.85} \times L_{37}^{-0.71} \mathrm{~s} ,
\end{equation}
$\alpha$ is $\sim$ 0.023 from their best fit. We obtain a predicted QPO frequency of $\sim$1.2 mHz, which is smaller than our 40 mHz QPO. The magnetic disk precession model (MDPM) can be able to explain mHz QPO with frequencies of the order of 1 mHz in several systems such as 4U 1626-67 \citep{2002ApJ...565.1134S} and 4U 0115+63 \citep{2019ApJ...872...33R}.  

These mechanisms above may not explain the 40 mHz QPO features in Cen X-3. A traditional description that QPOs originate from the expected inner radius of NS has encountered difficulty. In accretion-powered pulsars, the accretion disc is disrupted at the magnetospheric boundary and the torque of accreted matter transferred to the neutron star can cause spinning up in the neutron stars like system Cen X-3. If the neutron star's rotation is faster than the Kepler motion of the accretion disc, the accreted materials on to the neutron star will be halted due to centrifugal forces. Hence, the neutron star spin period would remain in equilibrium and the equilibrium spin period can be described as 
\begin{equation}
P_{\mathrm{eq}} \sim 3 M^{-2 / 7} R_{6}^{-3 / 7} L_{37}^{-3 / 7} \mu_{30}^{6 / 7} \mathrm{~s}.
\end{equation}
Using the approximate estimation of the magnetic strength of the neutron star, the equilibrium spin period of Cen X-3 is  about 3 s, close to its spin period of $\sim$4.8 s. Thus Cen X-3 appears to be close to its equilibrium spin period.

We assume the possible scenario that the accretion disc is truncated by the strong magnetic field close to corotation radius $R_{\mathrm{co}}$ defined as
\begin{equation}
R_{\mathrm{co}}=\left(\frac{G M P_{\mathrm{spin}}} {4 \pi^{2} }\right)^{\frac{1}{3}}.
\end{equation}
The corotation radius is situated at $\sim$4 $\times$10$^8$ cm, at which the Keplerian frequency in the disc equals the spin frequency of neutron star. In this scenario, the instability that can lead to quasi-periodic oscillations flares arises at the inner edge of the truncated disc, where allows angular momentum exchange between the magnetic field and the disc by disc–field interaction. If the disc is truncated outside but near the $R_{\mathrm{co}}$,  accretion can be inhibited by centrifugal force, in which the angular momentum is transferred from the star to the disc. As gas piles up, the increased gas pressure will push the truncated disc inwards until it crosses inside $R_{\mathrm{co}}$. Inside the $R_{\mathrm{co}}$, once the gas is accreted on to the star, gas pressure decreases, and the inner edge of the disc moves back outside the $R_{\mathrm{co}}$. And then another cycle starts. \cite{2010MNRAS.406.1208D} suggest that the period of the cycle varies from 0.02 to 20 $t_{\mathrm{visc}}$ (the viscous time-scale of $\sim$ a few hundred seconds in Cen X-3) and depends on the mean accretion rate. Thus the QPO frequency drift over the orbital phases may be related to the varying accretion rate. This physical process can provide a good description to understand the observed mHz oscillations in Cen X-3. Hence, the accretion disc is probably to be truncated near the $R_{\mathrm{co}}$. \cite{2021MNRAS.503.6045D} also explain mHz QPOs by the multi-vortex model hypothesis \citep{PhysRevLett.100.174503} in an analogous condition.

Intriguingly, we found that, for the first time, QPO rms of Cen X-3 has a prominent energy dependence. As the energy increases, the QPO rms amplitude decreases. At higher energy (above 20 keV), there are no QPO features. It may imply that the oscillations features could be associated with the soft photon sources.The frequency drifts at different energy may be related to corresponding energy-dependent time delay.

The energy-dependent QPO time lag shows an overall soft delay in the range of 2-20 keV except time lag of the 5-10 keV band is positive (maybe due to the iron emission lines by the reflection processes) and a negative time lag $\sim -50$ ms above 10 keV. \cite{Lee_2001,2014MNRAS.445.2818K,2020MNRAS.492.1399K,2022MNRAS.tmp.1048P} explain possible soft lags by inverse Compton scatterings for the kilohertz quasi-periodic oscillations of the LMXB NS system (such as 4U 1608–52). In the simple Comptonization models \citep{1997ApJ...484..383C,1999ApJ...510..874N}, Compton upscattering occurs not in the disc, but in the corona where photons come from the soft seed photons of either the accretion disc, the surface of the neutron star, or both. One can naturally expect the hard time lag as the hard photons scatter more than the soft ones. The soft time lags can be produced when a significant fraction of these photons Comptonized in the corona return to the disc and re-heat it. \citep{2014MNRAS.445.2818K} describe the observed soft time-lags as well as the increasing rms versus energy very well in NS 4U 1608–52 for the case when oscillation in the heating rate of the corona produces the QPO. In our results, the decreasing rms with energy is not consistent with the possible scenario where the oscillation derives from the corona. And that, the big variations (tens of ms) of time-lag found for the $\sim$40 mHz QPO would require the expected corona size to be at least the order of $\sim$ 1000 km, which is around the corotation radius.

As mentioned above, a possible scenario is that the energy-dependent quasi-periodic oscillation time lags could be ascribable to the varying accretion rate, viscosity, or the interaction region between the disc and field when the inner edge of the truncated disc is close to the corotation radius. As an example, in an oscillation cycle that consists of alternating accreting and quiescent states, the hard components of energy spectrum could radiate early. However, the structure of the instability turbulence is still not clear, future hydrodynamic or magnetohydrodynamic (MHD) simulations may help to understand these mHz QPOs.

\section{Conclusion} \label{sec:conclusion}

In summary, we have detected $\sim$40 mHz oscillations in Cen X-3 from observations with Insight-HXMT in 2017-2020. These QPOs were detected in a very soft spectral state for two observations in 2020. We study the evolution of mHz QPO frequency and rms amplitude, respectively, and find that the QPO frequency is ($\sim 33 -39$ mHz) in the phase $0.1-0.4$, and increases to $\sim 37- 43$ mHz in the orbital phase $0.4-0.8$. The frequency has a no dependence on X-ray intensity. In energy-dependent QPO analysis, we first report an anti-correlation of rms amplitude with energy in Cen X-3; The corresponding energy-dependent QPO time-lag of 2-20 keV has similar behavior: positive time lag of $\sim 20 $ ms around 5 -- 10 keV, and positive time lag of $\sim -(20-70)$ ms from 10 -- 20 keV. We discuss the possible scenario of the oscillations that occur in the truncated disc near the corotation radius to understand the origin of mHz QPO in Cen X-3. This scenario would be expected to apply to other mHz quasi-periodic flares in neutron star accreting systems.

\section*{Acknowledgements}
We are grateful to the referee for the fruitful suggestions to improve the manuscript. This work is supported by the National Key Research and Development Program of China (Grants No. 2021YFA0718500, 2021YFA0718503), the NSFC (12133007, U1838103, U1838201, U1838202). This work has made use of data from the \textit{Insight}-HXMT mission, a project funded by China National Space Administration (CNSA) and the Chinese Academy of Sciences (CAS).

\section*{Data Availability}
Data that were used in this paper are from Institute of High Energy Physics Chinese Academy of Sciences(IHEP-CAS) and are publicly available for download from the \textit{Insight}-HXMT website. To process and fit the power spectrum, this research has made use of XRONOS and FTOOLS provided by NASA.

\bibliographystyle{mnras}
\input{main.bbl}

\label{lastpage}
\end{document}